\documentclass[sn-mathphys]{sn-jnl}

\jyear{2021}%

\theoremstyle{thmstyleone}%
\theoremstyle{thmstyletwo}%

\theoremstyle{thmstylethree}%

\usepackage{amsmath}
\usepackage{listings}
\usepackage{xcolor}

\begin{document}

\title[Article Title]{KAN-LSTM-Transformer Neural Networks, MFV and Cosmological Parameters}


\author*[1,2,3]{\fnm{Jiang} \sur{Zhang}}\email{zhangphysics@126.com}

\author*[1]{Yan-dong Chen} \email{383876806@qq.com}
\affil[1]{School of Basic Sciences, Tianjin Agricultural University, Tianjin 300384, China }
\affil[2]{Physics Experimental Center, Department of Mathematics and Physics, Hebei GEO University, Shijiazhuang 050016, China}
\affil[3]{Hebei Key Laboratory of Optoelectronic Information and Geo-detection Technology, Hebei
	GEO University, China}


\abstract{Reconstructing the cosmic distance ladder directly from observations is a crucial issue in cosmology. In this paper, we present a novel method for modeling the cosmic distance ladder and estimating cosmological parameters through the use of  Kolmogorov-Arnold networks (KAN), Long Short-Term Memory (LSTM),  and Transformer networks (collectively referred to as KLT-Net), based on the apparent magnitude data from the Pantheon SN Ia compilation. 
As a data‑driven, non‑parametric method for reconstructing the distance modulus $\mu(z)$,
KLT‑Net  is shown to be highly effective in capturing the intricate, nonlinear measurement distributions.
After validating against various statistical and machine learning models, we have identified it as the most effective choice among the considered alternatives and ablation experiments. 
Subsequently, the statistical inference of $H_0$ and $\Omega_{\rm m}$  adopts the flat $\Lambda$CDM framework.
Moreover, we introduce the Most Frequent Value (MFV) approach to evaluate the absolute magnitude, $M_B$, from existing literature data.
In addition, we employ the Hessian matrix to validate the Bayesian method, demonstrating that the Hubble constant can be precisely constrained from the KLT-Net predictions within the flat 
$\Lambda$CDM framework. 
The integration of KLT-Net, the MFV approach, and Bayesian statistics establishes a robust framework for inferring cosmological parameters. This methodology facilitates future cosmological research, particularly in the analysis of complex datasets and the exploration of high-dimensional parameter spaces.

}

\keywords{ Cosmology; Neural networks; Stellar distance; Type Ia supernovae; Cosmological parameters}



\maketitle

\section{Introduction}\label{sec1}

Modern astrophysics and cosmology are facing unprecedented data challenges. With the promotion of large-scale sky survey observation projects (such as the LSST, Euclid, and JWST), massive multi-band observation data pose a serious challenge to traditional analysis methods \cite{Zhan2018RPPh,Shah2024MN,Kumar2025PRD}. In cosmology, accurate reconstruction of the cosmic distance ladder, determination of the Hubble constant ($H_0$), and exploration of the nature of dark energy all depend on efficient modeling and analysis of complex observational data \cite{Mukherjee2024ApJ,Shah2024ApJS}. In recent years, deep learning technology has shown unique advantages in the fields of astrophysical parameter estimation \cite{Moya2022A&A663}, cosmological reconstruction \cite{Mukherjee2024JCAP}, and observational data modeling \cite{Dialektopoulos2024PDU,Fortunato2025JCAP}. However, the traditional neural network architectures still exhibit significant limitations in dealing with high-dimensional nonlinear relationships.

Understanding the precise distances of astronomical objects at various redshifts is essential for inferring the history of cosmic expansion \cite{Di_Valentino2025}. Nevertheless, from an observational perspective, this task is not simple, given the absence of a uniform distance metric that applies to all scales of interest in cosmology. 
Therefore, researchers have adopted the progressive technique of calibrating distances, known as the ``cosmic distance ladder" approach, utilizing the intersecting regions of potentially diverse standardizable entities as ``rungs". The traditional distance ladder method commences with direct measurements of geometric distances, gradually calibrating Cepheid variables or the tip of the red giant branch stars, and ultimately calibrating Type Ia supernovae (SN Ia) \cite{Riess2024ApJ977,Riess2024ApJ962,Madore2024,Freedman2020,Freedman2023}.
By contrast, the inverse distance ladder commences with cosmological constraints from the drag epoch of the Cosmic Microwave Background (CMB) to the sound horizon, then proceeds to calibrate distances to Baryon Acoustic Oscillations (BAOs), and finally extends to SN Ia at lower redshifts \cite{Camarena2020MN,Gong2024,Adame2025JCAP}. SN Ia stand out as the primary endpoint for both approaches due to their consistent reliability as standard candles across a wide range of redshifts \cite{Riess2007ApJ,Scolnic2018}.

A cosmological model, such as the Lambda cold dark matter ($\Lambda$CDM) model, provides a physical theory for the expansion history of a spatially flat, homogeneous, and isotropic universe. The $\Lambda$CDM model is assumed to be applicable across all observed scales, from the present epoch to the epoch of recombination, and is currently the standard model with six free parameters determined by measurements.
Precise calculation of likelihood functions is a fundamental requirement in traditional parameter estimation methodologies.
Utilizing astronomical observation data and theoretical models, Bayesian inference is a prevalent approach for inferring the posterior distribution of model parameters. 
The growing intricacy of observational data and the high dimensionality of parameter spaces frequently make likelihood functions computationally intensive or even difficult to resolve \cite{Wang2023ApJS,Moriwaki2023RPPh}.
In order to address these constraints, a novel neural network-driven approach for likelihood-free inference has been developed recently \cite{Wang2023ApJS}.

There are numerous time series features present in astronomical observation data.
Although Gaussian processes (GPs) excel in handling this task by allowing for analytical predictions of derivatives and errors, they perform poorly in regions with sparse data and almost entirely fail in the absence of any data.
In time series prediction and sequence modeling, Long Short-Term Memory (LSTM) stands out as a variant of recurrent neural networks (RNNs) that effectively captures long-term dependencies in sequential data through gate mechanisms (input gate, forget gate, output gate).
The recent study by Shah et al. (2024) introduced the LADDER method, which utilizes an LSTM architecture in conjunction with covariance information to achieve model-independent reconstruction of the cosmic distance ladder using the Pantheon supernova dataset \cite{Shah2024ApJS}.

In contrast, the LADDER method significantly reduces reconstruction errors within the redshift range compared to GPs, yet the estimation of higher-order derivatives remains subject to substantial uncertainties that cannot be ignored \cite{Dialektopoulos2024PDU}. 
Similarly, Moya et al. found that as the input parameter dimension increased from low to high, the mean squared error of the traditional Multilayer Perceptrons (MLPs) nonlinearly increased when estimating stellar mass and radius via a stacked model \cite{Moya2022A&A663}. This indicates that existing architectures lack sufficient modeling capability for high-dimensional nonlinear relationships.

Furthermore, the Kolmogorov–Arnold network (KAN) surpasses traditional MLPs in parameter efficiency and interpretability by decomposing multivariable functions into combinations of univariable functions based on the Kolmogorov–Arnold Representation Theorem. It particularly excels in approximating high-dimensional nonlinear functions \cite{Liu2024arXiv240419756}. Additionally, by leveraging a self-attention mechanism, the Transformer achieves global dependency modeling through dynamically weighting the relationships, and eliminates the recursive nature of recurrent computations. This capability allows for parallel processing of lengthy sequential data, leading to a notable improvement in the extraction of intricate patterns.

A good model should have the following advantages: robust extrapolation capability ---  maintaining stable predictions  in data-sparse regions (e.g., high-redshift domains) via techniques like ensemble learning or positional encoding enhancements; 
effective covariance modeling --- integrating covariance information between data points through Monte Carlo sampling or shrinkage estimators to improve generalization; and rigorous validation --- demonstrating the superiority of architectures like LSTM over traditional models such as MLPs through ablation experiments.

Nevertheless, typical LSTM models exhibit key limitations: local dependency constraints --- while adept at capturing local sequence features, they are inadequate in capturing global evolutionary patterns across cosmic distance scales; parameter redundancy --- traditional LSTM layers contain numerous parameters, which can potentially impair training efficiency and weaken model generalization; and the long-range dependency challenge --- particularly when extrapolating to extremely high redshifts, gradient propagation in LSTM may struggle with limitations because of vanishing gradients or memory saturation over cosmological scales.

In recent years, network architectures incorporating the Transformer technique \cite{Vaswani2017}, such as Trans-UNet, have also been applied to image segmentation problems. For example, Galaxy–Galaxy strong lensing with U-Net (GGSL-UNet) can extract two-dimensional information from multiband images in both ground- and space-based observations \cite{Zhong2025ApJS}.
Moreover,  researchers \cite{Rozanski2025ApJ} recently proposed TransformerPayne and investigated the application of Transformer models for capturing long-range information in spectra, evaluating their performance in comparison to the Payne emulator.
However, the Transformer architecture is extremely complex, and is prone to overfitting when dealing with limited datasets. The recent advent of the Kolmogorov-Arnold network (KAN) offers a way to demystify the black-box nature of traditional neural networks, which has motivated researchers to integrate the KAN model into LSTM networks \cite{Cui2025arXiv250400392C}.

\begin{figure}
	\centering
	\includegraphics[width=\textwidth, angle=0]{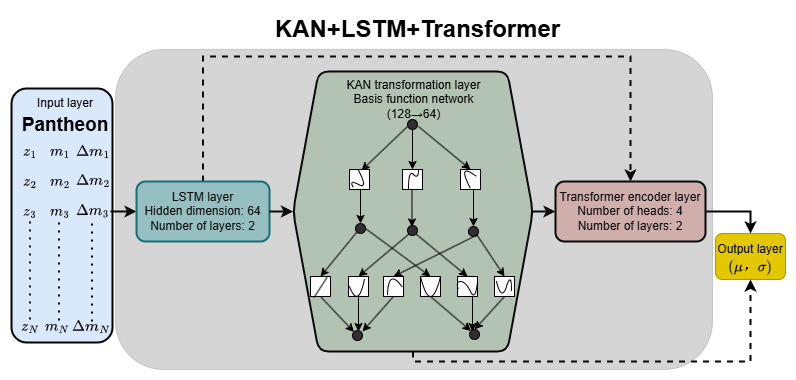}
	\caption{Conceptual diagram for training algorithm. }
	\label{Fig1}
\end{figure}

In response to the above issues, this paper proposes the KAN-LSTM-Transformer networks (KLT-Net), which achieves a breakthrough by integrating LSTM, KAN, and Transformer in a hybrid architecture as shown in Figure \ref{Fig1}.
The model operates in three key stages: First, the LSTM layer captures local temporal dependencies in sequences, inheriting the covariance modeling advantage demonstrated in approaches such as LADDER. Second, the KAN layer nonlinearly transforms features to approximate complex functions with fewer parameters, thereby enhancing computational efficiency and interpretability. Finally, the Transformer encoding layer extracts global contextual information through its self-attention mechanism, significantly enhancing the overall pattern recognition within the cosmic inflation history.

In this work, we utilize KLT-Net to investigate the cosmic distance ladder based on the Pantheon SN Ia dataset, considering the associated errors and full covariance information of the data. The paper is structured in the following way: In Section II, the data and model of this work are described. Section III presents the calculated results and reconstructs functions of $H(z)$ using the KLT-Net technique, along with a description of the simulations. Finally, discussion and conclusions are provided in Section IV.

\section{Data and Methodology}
\label{sect:data}
\subsection{Data}
In this study, our goal is the model‑independent reconstruction of the distance modulus $\mu(z)$ from Type~Ia supernova data. Following the prescription mentioned in the introduction, we use the Pantheon+ compilation of SN Ia measurements for data-driven machine learning applications \cite{Scolnic2018,Scolnic2022}. 
This dataset comprises 1701 light curves of 1550 unique, spectroscopically confirmed SN Ia, covering a wide redshift range  from 0.01 to 2.3, with more data available at lower redshifts and fewer data points at higher redshifts.
As this dataset provides a robust foundation for the model‑independent reconstruction of the distance modulus $\mu(z)$ \cite{Shah2024ApJS}, it plays a crucial role in analyzing the equation of state of dark energy within the distance-ladder investigation.

Furthermore, the Pantheon+ compilation contains apparent magnitude measurements ($m$) and their associated statistical errors ($\Delta m$), organized by various redshifts. 
In addition, the covariance matrix $C_{\text{sys}}$ of this dataset indicates the systematic errors and correlations within the measurements. Hence, the Pantheon+ compilation enables a comprehensive examination of cosmic distances across a broad range of redshifts, making it ideal for analyses that do not rely on specific models.

According to the classical cosmological model, the spatial part of the classical spacetime exhibits uniformity and isotropy in all directions, corresponding to Friedmann–Lemaître–Robertson–Walker (FLRW) cosmos \cite{Dodelson2020book}. The luminosity distance in  a flat FLRW cosmos is expressed by the following formula:
\begin{equation}
	\label{eq1}
	D_L(z)=c(1+z)\int^{z}_0 \frac{d\widetilde{z}}{H(\widetilde{z})}.
\end{equation}
This is associated with the Hubble parameter $H(z)$ at a given redshift $z$ \cite{Mukherjee2024JCAP}. With knowledge of the distance modulus, it is possible to calculate the luminosity distance $D_L(z)$ without relying on any cosmological model, as follows:
\begin{equation}
	\label{eq2}
	\mu_L(z)=m(z)-M_B=5{\rm log}_{10} \frac{D_L(z)}{\rm Mpc}+25.
\end{equation}
Recent studies have also revealed that the absolute magnitude $M_B$ of supernovae may exhibit characteristics of redshift evolution \cite{Camarena2021MN}.
Similar to the analysis by Shah et al. \cite{Shah2024ApJS}, subsequently, we also propagate the uncertainties in $z$ to $m(z)$, associated with the related covariance matrix $C_{\text{sys}}$ and statistical uncertainties $\Delta m$.

Using the Pantheon+ compilation, our objective is to train a KLT-Net that can learn effectively and extrapolate to higher redshifts, independent of the specific cosmological model, based on the apparent magnitude measurements. Figure \ref{Fig1} shows the schematic overview of the training algorithm for the KLT-Net model.

\begin{figure}
	\centering
	\includegraphics[width=\textwidth, angle=0]{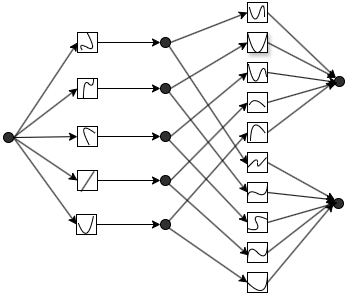}
	\caption{The structure of the KAN model.}
	\label{KAN}
\end{figure}

\subsection{Methodology}
\subsubsection{Kolmogorov–Arnold Network}
From the perspective of statistical machine learning, concerns have been brought to light in recent investigations using the Pantheon+ dataset. 
In this work of cosmic distance ladder reconstruction, we aimed to extend the LADDER framework developed by Shah et al. \cite{Shah2024ApJS}, which integrates the full covariance information of the Pantheon supernova dataset via the KLT-Net technique, in order to realize model-independent distance modulus reconstruction. 
Based on the Kolmogorov-Arnold representation theorem, KAN is a unique neural network architecture, with Figure \ref{KAN} illustrating the structure of the KAN neural network. 
We will discuss the methodologies used in this study, including the KAN, LSTM networks and the Transformer model, briefly explaining  their theoretical foundations and practical background.

The Kolmogorov-Arnold representation theorem affirms that any continuous high-dimensional function can be represented as a finite combination, via additive operations, of univariate continuous functions, as follows:
\begin{equation}\label{eq:KAT}
	f(x) = f(x_1,\cdots,x_n)=\sum_{q=1}^{2n+1} \Phi_q\left(\sum_{p=1}^n\phi_{q,p}(x_p)\right),
\end{equation}
where $\phi_{q,p}:[0,1]\to\mathbb{R}$ and $\Phi_q:\mathbb{R}\to\mathbb{R}$. This equation demonstrates that a multivariable function can be essentially reduced to a combination of appropriately defined univariable functions, with the combination involving only simple addition. 

Taking inspiration from the Kolmogorov-Arnold representation theorem, Liu et al. \cite{Liu2024arXiv240810205,Liu2024arXiv240419756} proposed the KAN as a promising alternative to MLPs. In contrast to traditional MLPs, the KAN completely eliminates linear weights, replacing each weight parameter with a univariate function parameterized as a learnable spline. 
This seemingly simple change makes the KAN superior to MLPs in terms of accuracy and interpretability, allowing it to achieve comparable accuracy with fewer neurons than MLPs, while also having fewer parameters.

While traditional MLPs conduct nonlinear spatial transformations on every layer, KAN specifically applies nonlinear transformations to each pair of bases before amalgamating them into a multidimensional space.
When depicted as a neural network graph, it acts like a two-layer neural network but lacks linear combinations. It directly applies nonlinear activation to input features, with the flexibility for these activation functions to be self-learned.
Essentially, KAN proves that a high-dimensional function can be broken down into learning a finite set of one-dimensional functions of polynomial complexity. Therefore, a KAN layer can be described as a matrix of learnable 1D functions:
\begin{align}
	{\mathbf\Phi}=\{\phi_{q,p}\},\qquad p=1,2,\cdots,n_{\rm in},\qquad q=1,2\cdots,n_{\rm out},
\end{align}
where $n_{\rm in}$ and $n_{\rm out}$ indicate the number of input features and the number of output features generated by a specific layer, respectively.
By stacking more KAN layers, the concept of deeper Kolmogorov-Arnold representations becomes more apparent, as shown in Figure 2.

From a statistical learning perspective, KAN can be seen as a fusion of spline functions and MLPs, leveraging their advantages and circumventing their drawbacks. In particular, each function $\phi_{q,p}$ can be expressed as a B-spline function, which is both exact and straightforward to fine-tune locally for low-dimensional functions. B-spline interpolation is a versatile and robust data fitting tool that operates by manipulating control points and weighting B-spline basis functions. The activation function $\phi_{l,j,i}$ in the matrix is a spline function that can be learned, and it can be represented as \cite{Bodner2024arXiv240613155}:
\begin{align}
	{\rm spline(x)}=\sum_ic_i B_i(x),
\end{align}
where $c_i$  denotes trainable coefficients. Due to its adjustable parameters through learning,  the B-spline function can effectively adapt to various complex function shapes. This grants a significant level of flexibility for the interaction between input features in network modeling.

Following the original parameter settings of Liu et al.  \cite{Liu2024arXiv240810205,Liu2024arXiv240419756}, the pre-activation of $\phi_{l,j,i}$ is simply $x_{l,i}$; the post-activation of $\phi_{l,j,i}$ is expressed by $\tilde{x}_{l,j,i}\equiv \phi_{l,j,i}(x_{l,i})$. The activation value of the $(l+1,j)$ neuron is simply the sum of all incoming post-activations: 
\begin{equation}\label{eq:kanforward}
	x_{l+1,j} =  \sum_{i=1}^{n_l} \tilde{x}_{l,j,i} = \sum_{i=1}^{n_l}\phi_{l,j,i}(x_{l,i}), \qquad j=1,\cdots,n_{l+1}.
\end{equation}
The matrix form is
\begin{equation}\label{eq:kanforwardmatrix}
	x_{l+1} = 
	\underbrace{\begin{pmatrix}
			\phi_{l,1,1}(\cdot) & \phi_{l,1,2}(\cdot) & \cdots & \phi_{l,1,n_{l}}(\cdot) \\
			\phi_{l,2,1}(\cdot) & \phi_{l,2,2}(\cdot) & \cdots & \phi_{l,2,n_{l}}(\cdot) \\
			\vdots & \vdots & & \vdots \\
			\phi_{l,n_{l+1},1}(\cdot) & \phi_{l,n_{l+1},2}(\cdot) & \cdots & \phi_{l,n_{l+1},n_{l}}(\cdot) \\
	\end{pmatrix}}_{{\Phi}_l}
	x_{l},
\end{equation}
where ${\mathbf \Phi}_l$ is the function matrix associated with the $l^{\rm th}$ KAN layer.
In a standard KAN network comprising $L$ layers, the output is obtained by providing an input vector ${x}_0\in\mathbb{R}^{n_0}$, given as:
\begin{equation}\label{eq:KAN_forward}
	{\rm KAN}({x}) = ({\Phi}_{L-1}\circ {\Phi}_{L-2}\circ\cdots\circ{\Phi}_{1}\circ{\Phi}_{0}){x}.
\end{equation}

\subsubsection{LSTM}
\label{sect:analysis}
In this part, we analyze the theoretical foundations of the LSTM methods in this work. 
For traditional neural network models, the inputs and outputs are typically assumed to be independent. Nevertheless, in real-world scenarios, there is often a certain relationship between them. RNNs are a common nonlinear dynamic model used in machine learning to handle such problems. Furthermore, RNNs exhibit superior performance on time series data when compared to the backpropagation algorithm and other artificial neural networks (ANNs). Yet, these sophisticated neural network configurations pose their own challenges. As supervised learning is taken into account and back-propagation is utilized for updating the weights of neural network connections, gradient vanishing and exploding emerge as major hurdles to the successfully training RNNs \cite{Lipton2015RNN,Das2023RNN}.
When RNNs are used to update the neural network's connection weights through backpropagation, they may struggle to update the influence of distant data on the weights if the current output is related to a distant time series. Hence, the introduction of LSTM is necessary to solve this limitation \cite{Van2020reviewLSTM,Mienye2024RNN}.

The LSTM model \cite{Hochreiter1997LSTM}, a classic recurrent neural network (RNN) architecture, is extensively utilized in sequence modeling and time series data processing tasks. When compared to traditional RNN models, LSTM models are better equipped to tackle the problems of gradient vanishing and exploding in long sequence data, while also being able to capture long-term dependencies in time series data \cite{Yu2019reviewLSTM}. Consequently, LSTM neural networks are better equipped for long-term storage and retrieval of information than traditional recurrent neural networks, especially when the network is unable to make sense of recent information.

\begin{figure}
	\centering
	\includegraphics[width=\textwidth, angle=0]{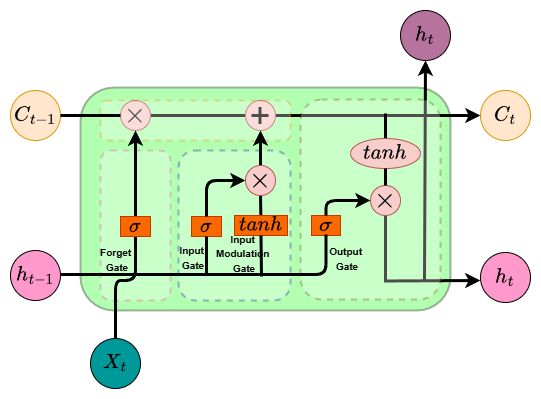}
	\caption{LSTM mechanism module.}
	\label{LSTM}
\end{figure}

The LSTM network comprises an input layer, one or more hidden layers, and an output layer. In comparison to RNNs, LSTM networks have their main characteristics embedded in the hidden layer, which consists of structures known as memory cells \cite{Mienye2024RNN,Shiri2023reviewLSTM}. 
The crucial aspect of LSTM networks is the control of the cell state, achieved by utilizing three control gates within the storage unit: the forget gate $g_t^f$, the input gate $g_t^i$, and the output gate $g_t^o$. Meanwhile, these gates are responsible for storing and adjusting the cell state. The forget gate determines how to retain the cell state, the input gate determines how to incorporate the current state into the cell state, and the output gate determines whether to use the cell state as the current output of the LSTM \cite{Hochreiter1997LSTM,Gers2000LSTM,Graves2013LSTM}. Figure \ref{LSTM} depicts the status update process in the inner structure of the LSTM unit. The status update mechanism within the LSTM layers is represented by the following equations: 
\begin{eqnarray}
	g_t^i & =& \text{sigmoid} \left( W_i \cdot [h_{t-1}, x_t] + b_i \right),\\
	g_t^f &=& \text{sigmoid} \left( W_f \cdot [h_{t-1}, x_t] + b_f \right),\\
	g_t^o &=& \text{sigmoid} \left( W_o \cdot [h_{t-1}, x_t] + b_o \right),\\
	\tilde{C}_t& =& \tanh \left( W_C \cdot [h_{t-1}, x_t] + b_C \right),\\
	C_t &= & g_t^f \bigotimes C_{t-1} + g_t^i \bigotimes \tilde{C}_t,\\
	h_t &=& g_t^o \bigotimes \tanh(C_t).
\end{eqnarray}
In the above mathematical formulations,   $W_\ast$ is the weight matrix for learnable parameters and $b_\ast$ is the bias term, where * is utilized instead of $f$ , $i$, $o$, or $c$ to signify the specified gates and memory cell state.
$C_t$,$\tilde{C}_t$,$h_t$ and $x_t$ stand for the memory cell state, candidate cell state, hidden state, and input data at time $t$, respectively. The symbol $\bigotimes$ denotes the component-wise operation.

\begin{figure}
	\centering
	\includegraphics[width=\textwidth, angle=0]{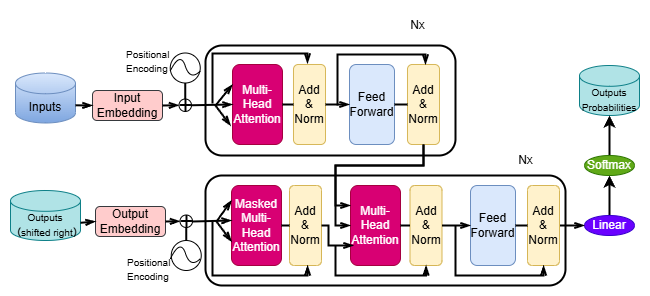}
	\caption{Schematic overview based on Transformer.}
	\label{Trans}
\end{figure}

\subsubsection{Transformer}
When facing complex features in astronomical data, it is difficult to pay more attention to intricate features, but attention mechanisms can focus on specific information and ignore irrelevant information. The Transformer is a neural network model based on the self-attention mechanism, which has a powerful expressive capability for effectively extracting global information \cite{Vaswani2017,Kitaev2020reformer}, as shown in Figure \ref{Trans}.

Compared to traditional RNNs and CNNs, the Transformer model excels in handling long sequential data and capturing long-range dependencies. The core idea of the Transformer model is the self-attention mechanism, which can establish relationships between different positions in a sequence and use this information for feature representation and context understanding.
By utilizing the self-attention mechanism, the Transformer model is able to capture dependencies between positions in a sequence.
As a result, it possesses strong parallelization abilities without relying on conventional recursion and convolution operations, drastically accelerating the training process \cite{Devlin2019Transformer}. Currently, the Transformer model has become a primary model in natural language processing (NLP) and other sequence-related tasks such as machine translation, text generation, and language understanding. 

The Transformer is a model based on the self-attention mechanism, typically consisting of multi-head self-attention (MSA) and a feed-forward network (FFN). The diagram in Figure \ref{Trans} illustrates the structure of the Transformer. 
Within the original encoding-decoding framework, the encoder layer comprises MSA and FFN, whereas the decoder layer incorporates an encoder-decoder attention mechanism.
By measuring the similarity between queries and keys, the self-attention mechanism determines the weight distribution of values and calculates attention coefficients for each sequence with respect to other sequences, capturing internal correlations between data or features. To begin with, it uses parameter matrices $W_q$, $W_k$, $W_v$ to linearly transform the input matrix X, resulting in projection matrices Q, K, V for X on query, key and value, with dimensions $d$. 
Subsequently, the self-attention weight matrix $QK^T$ is computed by interacting through the dot product of the matrix for similarity calculation. To avoid gradient vanishing, the $QK^T$ is divided by a coefficient ($\sqrt{d_k}$) and then normalized to obtain attention weights, via the softmax function to obtain a probability distribution where the weights sum up to 1.

By evaluating the attention weight distribution to assess the importance of information at different positions, the computation is carried out with the matrices to generate a refined value of combined attention, ultimately enhancing the network's proficiency in recognizing essential features. To summarize, the formula is as follows:
\begin{eqnarray}
	\text{head}=\text{Attention}(Q,K,V)={\rm softmax}(\frac{QK^T}{\sqrt{d_k}})V.
\end{eqnarray}

The multi-head self-attention mechanism consists of multiple parallel self-attention layers. First, the input matrix X is transformed and passed through h different self-attention layers to calculate the attention weight matrix $head_i$ via the scaled dot-product interaction operation.  After that, the h output matrices are merged together to produce a high-dimensional matrix, consolidating information from different feature dimensions. 
Lastly, the result is fed into a linear transformation layer, compressing the high-dimensional information of the output matrix by multiplication with a weight matrix $W^O$ to achieve the output matrix with the same dimensions as the input matrix. The mechanism of multi-head self-attention can be expressed as:
\begin{eqnarray}
	\text{MultiHead}(Q,K,V)=\text{Concat}(\{\text{head}_i\}_{i=1}^{i=h})W^O,
\end{eqnarray}
where h denotes the number of heads in multi-head self-attention.

The functionality of the encoder layer is achieved through the feed-forward layer, which consists of two linear transformations connected by a ReLU activation function in between them. Each position-wise component of the input sequence is processed independently by the network, maintaining consistent input and output dimensions. The formula for the linear transformation is given as follows \cite{Vaswani2017}:
\begin{eqnarray}
	\text{FFN}(x) = \max(0, xW_1 + b_1)W_2 + b_2.
\end{eqnarray}

\begin{figure}
	\centering
	\includegraphics[width=\textwidth, angle=0]{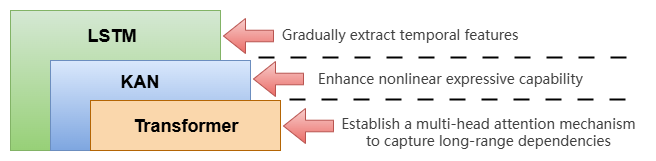}
	\caption{Schematic diagram of KAN, LSTM and Transformer.}
	\label{KLT}
\end{figure}

\subsubsection{KLT-Net}
The efficiency of the traditional LSTM architecture in capturing the evolving features of astronomical time series data may limit the statistical significance of conclusions. Some instances have already brought to light the major shortcomings of current deep learning methods in complex astrophysical problems: for example, their restricted nonlinear modeling capability --- the static forms of conventional activation functions (like ReLU, GELU) cause networks to be unable to autonomously adjust the combination of base functions. Moreover, a challenge arises from the inefficiency of feature interactions: LSTM's ability to model sequences makes it difficult to capture long-range dependencies across various redshift scales \cite{Clevert2015fast,Vaswani2017}. Additionally, it is important to take into account the quantification of uncertainty bias: simple variance predictions are unable to accurately model the complex distribution properties of heteroscedastic noise.

According to the theory above, we present the KLT-Net procedure, as illustrated in Figure \ref{KLT}. This module integrates the benefits of KAN, LSTM, and Transformer mechanisms. The main idea behind KAN is to employ spline functions for parameterized learning, which enhances nonlinear representation capabilities with fewer parameters and nodes, resulting in higher accuracy.
With its remarkable parallel computing capabilities and the capacity to capture complex dependencies, the Transformer can effectively handle large-scale data and complex sequence modeling. 
The LSTM neural network solves the problems of gradient explosion and gradient disappearance in RNNs by improving the RNN model, and achieves excellent performance in forecasting astronomical time series data. 
Further details on the LSTM technique can be found in the references \cite{Shah2024ApJS}.

The implementation of KLT-Net in this study includes the following key technologies: Basis function design: The cubic B-spline basis function is adopted; Adaptive mesh refinement: By introducing learnable node position parameters, dynamic optimization of basis function distribution is achieved; Parameter efficiency optimization: Through the weight sharing mechanism, the number of basis function parameters is reduced. 

It is important that all experimental models are assessed using consistent performance metrics, such as mean squared error (MSE), root mean square error (RMSE), and mean absolute error (MAE). Similar to the LADDER procedure \cite{Shah2024ApJS}, we also use MSE to evaluate the prediction performance by measuring the difference between predicted values and actual observations, with the standard formulas:
\begin{eqnarray}
	\text{RMSE}=\sqrt{\frac{\sum(y_i-\hat{y}_i)^2}{n}},
\end{eqnarray}
\begin{eqnarray}
	\text{MAE} =\frac{\sum \mid y_i - \hat{y}_i \mid}{n},
\end{eqnarray}
where $y_i$ represents the observed value, and $\hat{y}_i$ indicates the predicted value.

\subsection{Regression models based on statistical learning}
From a statistical learning perspective, evaluating cosmological parameters  can essentially be  viewed as a regression task. Historically, this issue has been resolved using cosmological models that take the observability of stars as input parameters for estimation. We recommend addressing this problem from a different angle and verifying the theoretical model with a statistical learning approach. With the visibility of stars as input features, we utilize a statistical learning model to evaluate the optimal regression model for cosmological parameters.

According to statistical learning theory \cite{Vapnik1999,Hastie2009}, numerous regression methods have been applied to similar studies on astronomical observation data samples \cite{Kelly2007,Feigelson2012,Moya2022A&A663}. As a result, we do not restrict ourselves to a specific artificial intelligence model, but instead evaluate a selection of AI solutions that represent the current best models. The aim is to provide a comparative study to elucidate the most suitable statistical learning model for handling the regression of cosmological parameters. Next, we briefly outline each statistical learning regression method used in our experiment.

As a powerful yet simple procedure for astronomical data analysis, the decision tree regression (DTR) model used for statistical learning is a non-parametric supervised learning method. By analyzing data features, it can acquire knowledge of straightforward decision rules to make predictions about the target variable, and construct a decision tree by dividing the dataset into smaller subsets through recursion. At each node, a feature and a threshold are selected to split the dataset into different sections, ensuring that the variance of the target variable is minimized in the resulting subsets \cite{Breiman1984,Loh2011,Loh2014}.

Random forest (RF) is a powerful and widely used ensemble learning algorithm that can be applied to both classification and regression problems \cite{Breiman2001,Genuer2020,Louppe2014}. Multiple decision trees make up the ensemble model known as a random forest. Each decision tree provides a prediction, and the random forest averages the predictions of all decision trees to obtain the final prediction result. By combining the predictions of multiple decision trees, it enhances the model's ability to generalize and stabilize, thereby mitigating the issue of overfitting in single decision trees. For each sampled subset of the data, a decision tree is constructed by randomly selecting a subset of features at every node split. The tree is grown by recursively partitioning nodes to minimize the mean squared error or variance reduction (for regression tasks), until a stopping condition (e.g., maximum depth, minimum samples per node) is met.

Support Vector Machines (SVM, \cite{Cortes1995SVM,Noble2006SVM}) were originally designed for classification problems, but have been improved to also be used for regression analysis. Support Vector Machine Regression (SVR,  \cite{Drucker1996SVM,Smola2004SVR,Brereton2010SVM}) is an extension of SVM for regression problems.
The goal of SVR is to find a regression function that minimizes the error between predicted and actual values within a certain tolerance range, while also ensuring the complexity of the function is kept as low as possible to improve the  generalization ability of the models. 
In order to satisfy the error requirement while minimizing model complexity, this is typically achieved by minimizing $\frac{1}{2}\|\omega\|^2+C\Sigma(\xi_i+\xi_i^\star)$.
When dealing with nonlinear regression problems, SVR usually introduces the kernel function $k (x_i, x_j) = \phi(x_i)^T \phi(x_j)$. Through the kernel function, input data can be mapped to a high-dimensional feature space, and a linear regression function can be found in the high-dimensional space, so as to realize nonlinear regression \cite{Awad2015SVM}.

K-nearest neighbor (KNN) is another fundamental classification and regression algorithm, belonging to the field of data mining and analysis \cite{Guo2003knn,Steinbach2009knn,Song2017KNN}. 
The basic principle of KNN is that samples of the same category tend to be closer in the feature space. Thus, the class of each sample can be determined by the class of its nearest k neighbors.
Specifically, for a sample to be classified, the algorithm calculates its distance from all other samples in the feature space, and identifies the nearest k samples (that is, ``nearest neighbors"). Which neighbors are 'nearest' depends on the distance metric (e.g., Euclidean, Manhattan). If most of these k neighbors belong to a certain category, the sample will also be classified into this category, because similar features usually correspond to the same category. 
The output in regression analysis is usually the average (or weighted average) of the target values of the k nearest neighbors.

ANNs have a distinctive nonlinear adaptive information processing ability that sets them apart from traditional artificial intelligence methods. They have successfully addressed the shortcomings in intuition, particularly in unstructured information processing, and have found applications in pattern recognition, intelligent control, combinatorial optimization, and prediction.
The MLPs regression model is a supervised learning algorithm based on ANNs, which is mainly used to solve prediction problems  \cite{Pinkus1999MLP,Taud2017MLP,Yu2024kanMLP}.
Its core structure consists of an input layer, multiple hidden layers and an output layer, in which each hidden layer contains several neurons (e.g., dozens of neurons in a layer), and the neurons are connected by weights and adopt nonlinear activation functions (e.g., ReLU) to introduce complexity.
The model optimizes the weights by the back propagation algorithm (e.g., Stochastic Gradient Descent, SGD) to minimize the error (e.g., mean square error) between the predicted value and the real value.
In order to avoid the trouble of manually adjusting the learning rate, we dynamically adjust the learning rate: if the loss function no longer decreases during training, the learning rate will be automatically reduced.

\begin{figure}
	\centering
	\includegraphics[width=\textwidth, angle=0]{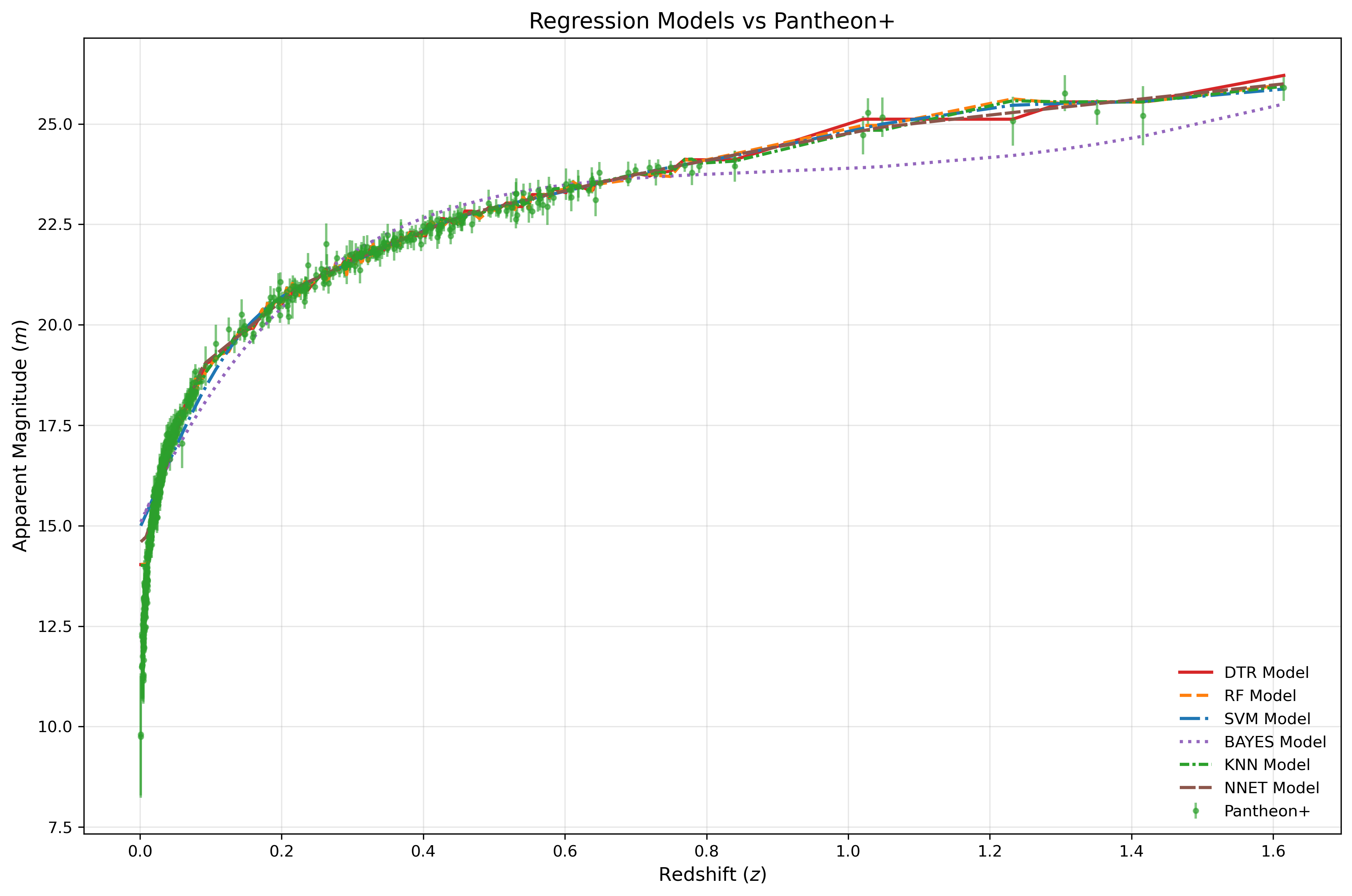}
	\caption{Comparison of different regression models with Pantheon+.}
	\label{multi-Model}
\end{figure}

As the number of iterations increases, the model converges on both the test and training sets, with minimal difference between them, ultimately reaching the preset target error. This indicates that the model's performance meets the evaluation standard. Figure \ref{multi-Model} displays the predictive capability of the MLP neural network (NNET) and other models after training on the measurement data. Following the previous work \cite{Moya2022A&A663,Shah2024ApJS,Ren2025JHyd},
MSE is used as the loss function of the MLPs model to evaluate the performance of the model, similar to RMSE and MAE shown in Table \ref{tab:model_results}. The core of MSE is to find the difference between the predicted result of the model and the actual target value, and to minimize MSE by adjusting the parameters of the model, so that the predicted value is as close as possible to the actual target value. 
The MSE value of the LADDER is 0.018495 \cite{Shah2024ApJS}, while the MSE value of KLT-Net is 0.015727.
The smaller the MSE value, the smaller the difference between the predicted result of the model and the actual target value, which indicates that the performance of the model is better.

\begin{table}[htbp]
	\centering
	\caption{Model Performance Comparison}
	\label{tab:model_results}
	\begin{tabular}{|l|r|r|r|}
		\hline
		\textbf{Model} & \textbf{MSE} & \textbf{RMSE} & \textbf{MAE} \\
		\hline
		DTR    & 0.038703 & 0.196730 & 0.144229 \\
		\hline
		RF     & 0.036460 & 0.190944 & 0.141995 \\
		\hline
		SVM    & 0.275372 & 0.524759 & 0.307892 \\
		\hline
		BAYES  & 0.488107 & 0.698646 & 0.529312 \\
		\hline
		KNN    & 0.037891 & 0.194655 & 0.140640 \\
		\hline
		NNET   & 0.062725 & 0.250449 & 0.189503 \\
		\hline
		KLT    & 0.015727 & 0.125409 & 0.098847 \\
		\hline
	\end{tabular}
\end{table}

\section{Ablation Study}
\label{sec:ablation}
We perform a comprehensive ablation study to assess the individual impact of each element within the KLT-Net model.
Ten independent random seeds (104, 35, 74, 108, 154, 185, 333, 335, 443, 514) are utilized for repeating all experiments, with subsequent reporting of the mean and sample standard deviation for all metrics. 
Following \citep{Shah2024ApJS}, the four complementary metrics are employed: MSE, RMSE, MAE, and the KL divergence ($D_{\rm KL}$). 
Quantifying the discrepancy between the predicted and reference distributions of the distance modulus, the latter captures both bias and uncertainty calibration.

Comparison is made among five model variants: \textbf{KLT} (full model; KAN + LSTM + Transformer), \textbf{LT} (without KAN), \textbf{KT} (without LSTM), \textbf{KL} (without Transformer), and the \textbf{LADDER} baseline, which is an LSTM-based architecture proposed by Shah et al. (2024) \citep{Shah2024ApJS}.

\begin{figure}
	\centering
	\includegraphics[width=\columnwidth]{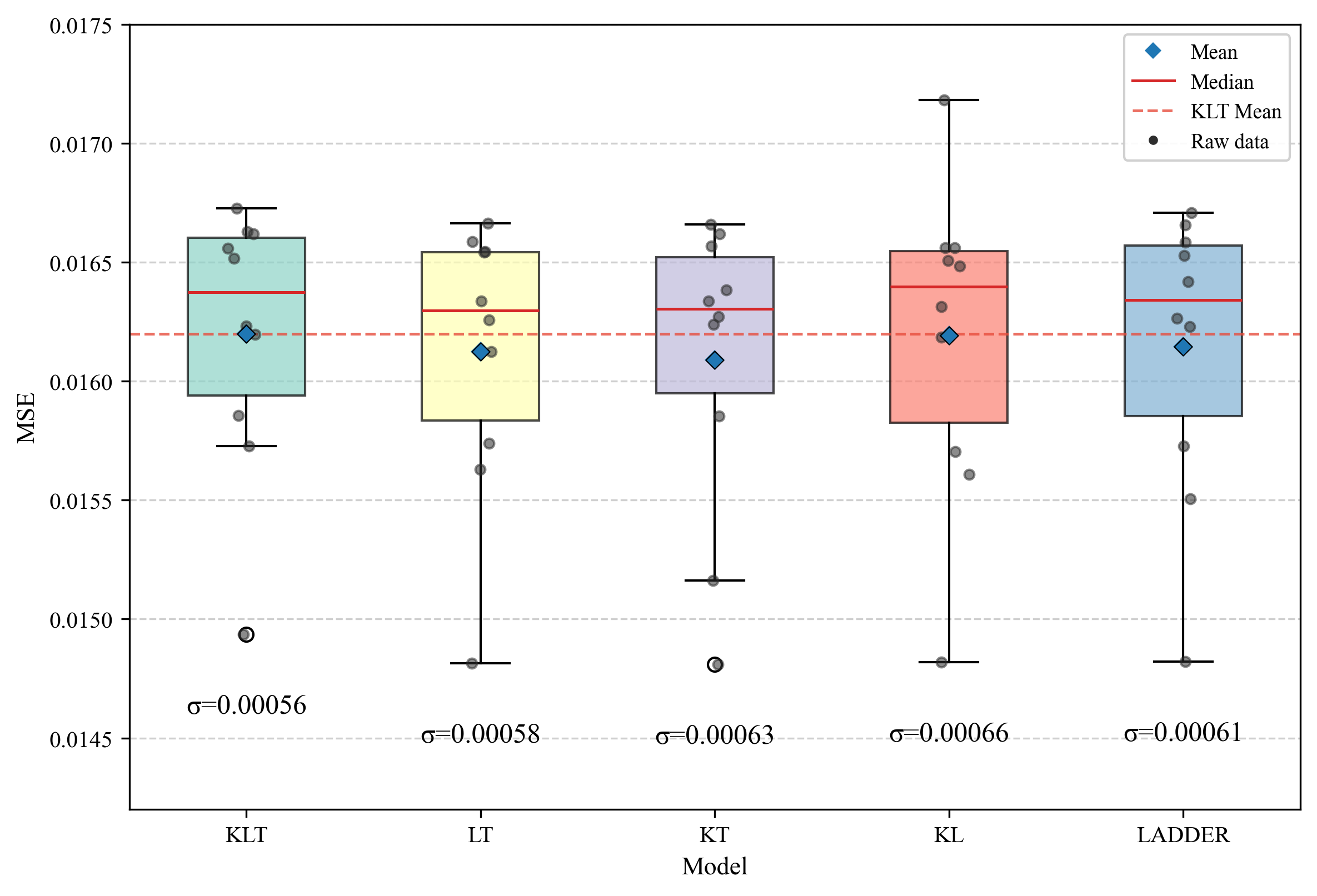}
	\caption{Boxplot of the MSE distribution across ten random seeds for five model variants.
		The central red line denotes the median, the blue diamond indicates the mean,
		and gray circles represent individual runs with horizontal jitter (Gaussian noise, $\sigma = 0.04$).
		The dashed red line marks the mean MSE of the full KLT‑Net.
		Numerical values above each box indicate the cross‑seed standard deviation $\sigma_{\rm MSE}$.
		KLT‑Net achieves the smallest variance ($\sigma_{\rm MSE} = 0.559\times10^{-3}$),
		highlighting its robustness to weight initialization.}
	\label{fig:ablation}
\end{figure}

Visualizing the MSE distribution across ten independent runs with different random seeds is Figure~\ref{fig:ablation}, with a summary of quantitative performance and stability metrics, including mean MSE, standard deviation, coefficient of variation (CV), and MSE range presented in Table~\ref{tab:ablation}.
The CV is defined as follows:
\begin{equation}
	\text{CV (\%)} = \frac{\mathrm{standard\ deviation}}{\mathrm{mean\ MSE}} \times 100\%.
\end{equation}
The CV metric removes the magnitude dependency of absolute errors, offering a reliable measure of relative training stability among various model setups.
With a mean MSE of $0.016089$, the KT variant stands out for achieving the optimal average prediction accuracy compared to all other model variants.
Despite this, the full KLT model displays the most consistent training stability, resulting in the smallest cross-seed error fluctuation, with $\sigma_{\rm MSE} = 0.559 \times 10^{-3}$ and $\sigma_{\rm RMSE} = 2.217 \times 10^{-3}$.
Additional quantitative stability verification affirms that KLT achieves the lowest CV of $3.45\%$ and the most limited MSE range, illustrating its superior robustness against random initialization.

For the purpose of statistically confirming the performance distinctions, paired t-tests and Wilcoxon signed-rank tests are executed for all model comparisons, as outlined in Table~\ref{tab:stat_test}.
The absence of a statistically significant advantage in mean reconstruction accuracy is evident across all models, with $p>0.05$ in all pairwise comparisons.
The marginal difference between KLT and LT ($p_{\rm t}=0.0582$, $p_{\rm w}=0.0645$) serves to reinforce that all models exhibit similar prediction accuracy across repeated training runs.

Above all, the exceptional stability of KLT is a crucial and sought-after characteristic for cosmological inference. 
By leveraging the synergistic integration of covariance-aware feature extraction, sequential modeling, and temporal enhancement modules, KLT achieves the most consistent reconstruction of the distance modulus $\mu(z)$ across various training sessions.
Utilizing the reconstructed $\mu(z)$ as the direct input for subsequent $\chi^2$ fitting and Bayesian MCMC sampling of $H_0$ and $\Omega_{\rm m}$ is crucial. This approach effectively reduces systematic and statistical biases in cosmological parameter estimation by suppressing random training fluctuations.

\begin{table*}
	\centering
	\caption{Ablation experiment results on the measurement dataset.
		Values are derived from MSE over ten random seeds.
		Mean MSE and standard deviation ($\sigma_{\rm MSE}$) are scaled by $10^{-3}$.
		CV denotes coefficient of variation (\%), and Range denotes the range of raw MSE values.
		The smallest mean MSE, standard deviation, and CV are highlighted in bold.}
	\label{tab:ablation}
	\begin{tabular}{lccccc}
		\hline\hline
		Model                  & KLT           & LT            & KT            & KL            & LADDER        \\
		\hline
		Mean MSE ($10^{-3}$)   & 16.199        & 16.123        & \textbf{16.089} & 16.192        & 16.144        \\
		$\sigma_{\rm MSE}$ ($10^{-3}$) & \textbf{0.559} & 0.580         & 0.632         & 0.659         & 0.610         \\
		CV (\%)                & \textbf{3.45} & 3.60          & 3.93          & 4.07          & 3.78          \\
		Range MSE ($10^{-3}$)  & \textbf{1.792} & 1.849         & 1.849         & 2.361         & 1.886         \\
		\hline
	\end{tabular}
\end{table*}

\begin{table*}
	\centering
	\caption{Statistical test results of paired comparisons.
		Significance level $\alpha=0.05$. ``No'' indicates no statistically significant difference between two models.}
	\label{tab:stat_test}
	\begin{tabular}{lccc}
		\hline\hline
		Comparison      & t-test $p$-value & Wilcoxon $p$-value & Significant? \\
		\hline
		KLT vs LADDER   & 0.4549           & 0.4316             & No            \\
		KLT vs LT       & 0.0582           & 0.0645             & No            \\
		KLT vs KT       & 0.1512           & 0.3750             & No            \\
		KLT vs KL       & 0.9168           & 0.2324             & No            \\
		\hline
	\end{tabular}
\end{table*}

When the KAN module (LT) is taken out, there is a rise in MSE variance, whereas the absence of the LSTM module (KT) reduces the mean MSE, albeit with slightly larger cross-seed fluctuations.
The Transformer module does not primarily aim to reduce point-wise MSE; rather, it serves as a regularizer to enhance training stability and refine global dependency modeling. 
By promoting smoother and more physically plausible reconstructions of the distance ladder, this behavior proves advantageous for downstream cosmological inference.

Each KLT variant attains competitive performance levels equivalent to LADDER. 
Paired statistical tests validate the lack of statistical significance in MSE differences ($p > 0.05$).
Nonetheless, KLT and its ablated forms consistently display decreased cross-seed variance, illustrating enhanced stability in comparison to the LSTM-based baseline.

\color{black}
\section{Cosmological Applications of KLT-Net}
\begin{figure}
	\centering
	\includegraphics[width=\textwidth, angle=0]{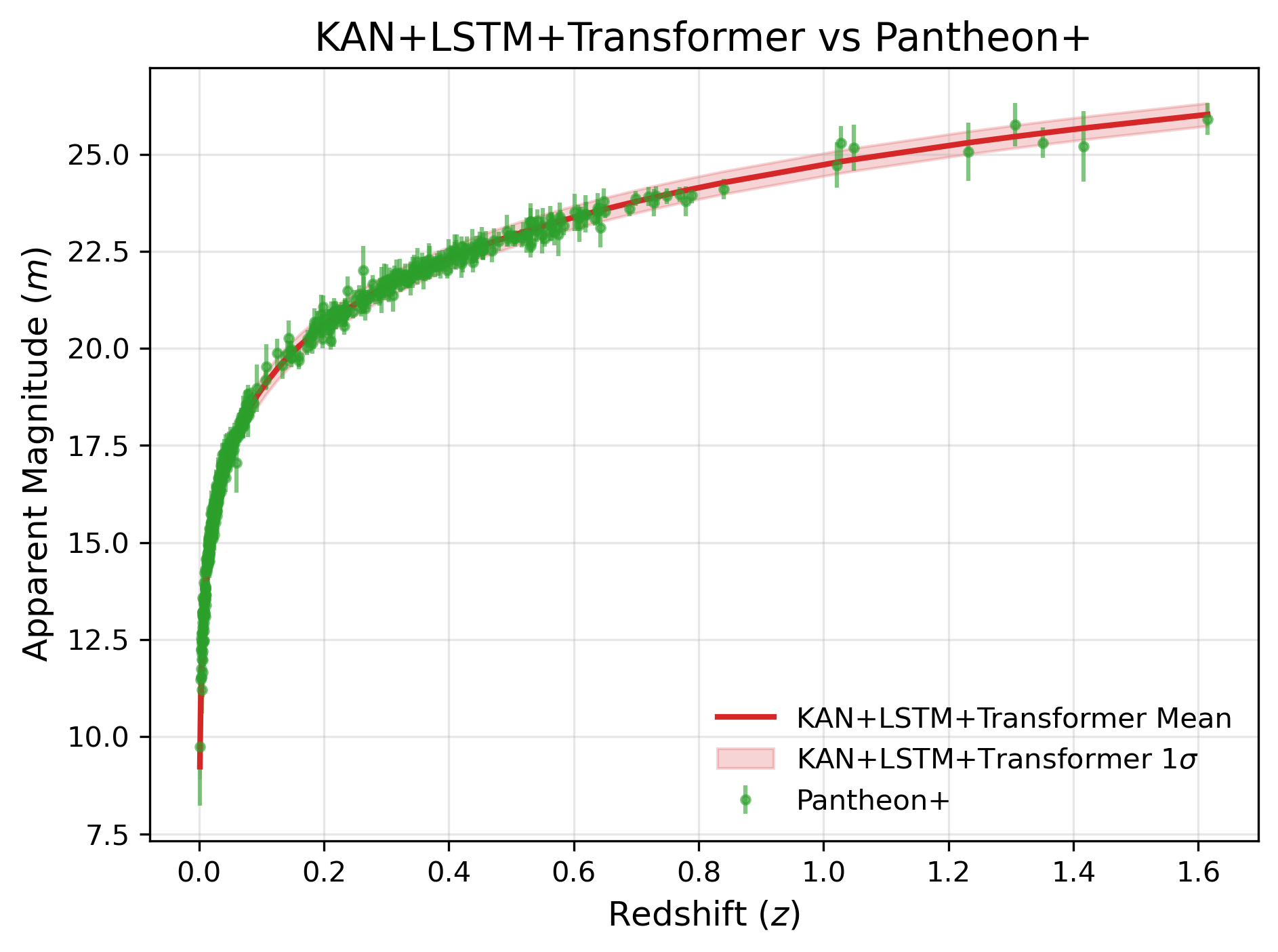}
	\caption{Comparison between Pantheon-trained KLT-Net reconstruction vs. Pantheon+.}
	\label{fit}
\end{figure}

After training KLT-Net with Pantheon, we can test the consistency of a similar SN Ia dataset in a model-independent way. Following the LADDER \cite{Shah2024ApJS}, we also select the Pantheon+ compilation \cite{Scolnic2022} as the most up-to-date publicly accessible SN Ia dataset at the time of writing. From Figure \ref{fit}, it can be seen that the KLT-Net model fits well with the Pantheon+ data. 
This holds potential importance for our comprehension of cosmic phenomena, and also indicates the possibility of inferring important parameters of astrophysics or cosmic processes from the KLT-Net model.
Only when the extended measurement dataset meets stringent reliability criteria can it significantly improve the accuracy of cosmological parameter estimation and enhance the robustness of fundamental physical models.

Over the last ten years, advancements in science and technology have led to better observation techniques, enabling us to gather more complete data to evaluate current theories in cosmology. Furthermore, observations of SN Ia confirm that the universe is undergoing accelerated expansion, which indicates that this expansion is being driven by dark energy. Parameters in cosmological models provide crucial details about the universe, including the degree of space curvature, density of matter, age of the universe, and rate of expansion. 
Determining the best-fitting values of these parameters is a key aspect of cosmological research, as it involves constraining the parameters of cosmological models using observational data.

The Hubble Law provides the primary observational evidence for the expansion of the universe. 
The $H_0$ is crucial in cosmology as it governs all absolute distances and time scales in the field, making the precise and certain determination of its value highly important. Since accurate determination of the Hubble constant plays a crucial role, the research on the $H_0$ is becoming more and more prominent. 

Recently, however, there has been an unbridgeable gap between the Hubble constant values measured by different measurement methods, the most prominent of which comes from two measurement approaches in the early universe and the late universe. The local direct measurements come from the late-time cosmic distance ladder measurements($H_0 \approx 73.2 \pm 0.9$ km s$^{-1}$Mpc$^{-1}$, \cite{Riess2022,Riess2024ApJ977}), while the global model fitting values come from the observational constraints of the cosmic microwave background radiation on the cosmological standard model($H_0 = 67.27 \pm 0.60$ km s$^{-1}$Mpc$^{-1}$, \cite{Planck2020}).
Clearly, there is a significant deviation of close to $5\sigma$ between them. If this deviation cannot be accounted for by the observational and/or systematic errors associated with these two measurement techniques, then it unquestionably presents a significant challenge to the prevailing cosmological standard model. This is the crisis of the Hubble constant \cite{Riess2020NatRP}.

The method of using the Hubble parameter as a means to directly measure the expansion of the universe has been receiving more and more attention in recent years. As a standard for testing cosmological models, the Hubble parameter not only helps determine important cosmological parameters affecting the evolution of the universe, but also can refine descriptions of key moments in the cosmic evolution process. In the $\Lambda$CDM model with a flat universe, according to the Einstein field equations, the Friedmann equations can be derived \cite{Liddle2015book,Dodelson2020book} as follows:
\begin{eqnarray}
	H^2=\frac{8\pi G}{3}\rho-\frac{k}{a^2}+\frac{\Lambda}{3},
\end{eqnarray}
where as an extra term, $\Lambda$ is the cosmological constant, and $\Omega_{\Lambda 0}=1-\Omega_{m0}$ based on $\Lambda CDM$. Following the derivation of classical cosmology \cite{Chantada2023PhRvD}, we can describe H(z) as
\begin{eqnarray}
	H(z)=H_0\sqrt{\Omega_{m0}(1+z)^3+1-\Omega_{m0}},
\end{eqnarray}
where $\Omega_{m0}$ is $\frac{8\pi G \rho _{m0}}{3H_0^2}$ and $H_0$ denotes the value of $H(z)$ at $z=0$.

\subsection{$M_B$ inferred via MFV}
In cosmology, it is very important to estimate the absolute magnitude $M_B$ of SN Ia, because the tension present in $H_0$ can also be seen as tension between the local and early universe constraints regarding the $M_B$ of SN Ia \cite{Mukherjee2024JCAP}.
There are three strategies for statistical central estimation of cosmological parameters $M_B$,  $H_0$ and $\Omega_{m0}$ in statistical learning: the mean, median, and MFV \cite{Steiner1988, Steiner1997, Szegedi2014, Szabo2018,Zhang2022,Golovko2023EPJC,Golovko2024}.  The MFV method was suggested to tackle the same issues \citep{Zhang2017, Zhang2018, Golovko2023Sensor, Zhang2024MN}. In order to understand the impact of prior distribution and error distribution, it is essential to utilize the MFV method in evaluating the characteristics of datasets. Unlike the maximum likelihood principle or Least Squares Method (LSM) \citep{Zhang2012CPL}, Steiner put forward the maximum reciprocals principle as follows:
\begin{eqnarray}
	\sum_{i} \frac{1}{X_i^2+S^2}=max,
\end{eqnarray}
where $X_i$ indicates the residuals and deviations, and $S$ is the measurement error. Furthermore, by iterating through the data, we can calculate both the most frequent value and the cohesion. In the (j+1)-th step of the MFV algorithm, the relative equation of iterations for the most frequent value $M$ is given as:
\begin{eqnarray}
	\label{eq1}
	M_{j+1}= \frac{ \sum_{i=1}^{n} \frac{\varepsilon_j^2x_i}{\varepsilon_j^2+(x_i-M_j)^2}} {\sum_{i=1}^{n}\frac{\varepsilon_j^2}{\varepsilon_j^2+(x_i-M_j)^2}},
\end{eqnarray}
where $x_i$ is a series of measurements and the dihesion $\varepsilon_j$ can be described by
\begin{eqnarray}
	\label{eq1}
	\varepsilon_{k+1}^2= \frac{3 \sum_{i=1}^{n} \frac{\varepsilon_k^4(x_i-M_j)^2}{[\varepsilon_k^2+(x_i-M_j)^2]^2}} {\sum_{i=1}^{n}\frac{\varepsilon_k^4}{[\varepsilon_k^2+(x_i-M_j)^2]^2}}.
\end{eqnarray}
At the start of the process, the initial value $M_0$ for the iteration is set as the arithmetic mean of the measurements, and the initial value of $\varepsilon$ is as follows:
\begin{eqnarray}
	\label{eq1}
	\varepsilon_0=\frac{\sqrt{3}}{2}(x_{max}-x_{min}).
\end{eqnarray}
The threshold criterion can be adjusted to control the precision in each iteration. Through a sequence of iterations, the most frequent value $M$ and dihesion $\varepsilon$  can be identified until the dihesion becomes less than a specified threshold value. 

The bootstrap method, as highlighted by Efron et al. \cite{Efron1994,Davison1997}, is indispensable for estimating the uncertainty of physical quantities and for determining the validity of the calculated results. 
The fundamental process for calculating confidence intervals using the bootstrap method is briefly described as follows.
Given that the dataset of $M_B$  consists of  ($M_{B_1}$,$\ldots$,$M_{B_i}$) selected from an independent and identical distribution of true values of $M_B$ with the corresponding statistic $\theta$($M_{B_1}$,$\ldots$,$M_{B_i}$), which is the MFV.
After generating a bootstrap sample ($M^{\textasteriskcentered}_{B_1}$,$\ldots$,$M^{\textasteriskcentered}_{B_i}$) from the initial data with replacement, the next step is to calculate the important statistic, MFV, for the bootstrap sample.
By repeating this process B times, typically between 1000 and 3000 times, the distribution of the MFV is produced.
Consequently, these distributions can help determine confidence intervals (typically 68.27 or 95.45 percent) of the MFV for different distance measurement techniques.

\begin{figure}
	\centering
	\includegraphics[width=\textwidth, angle=0]{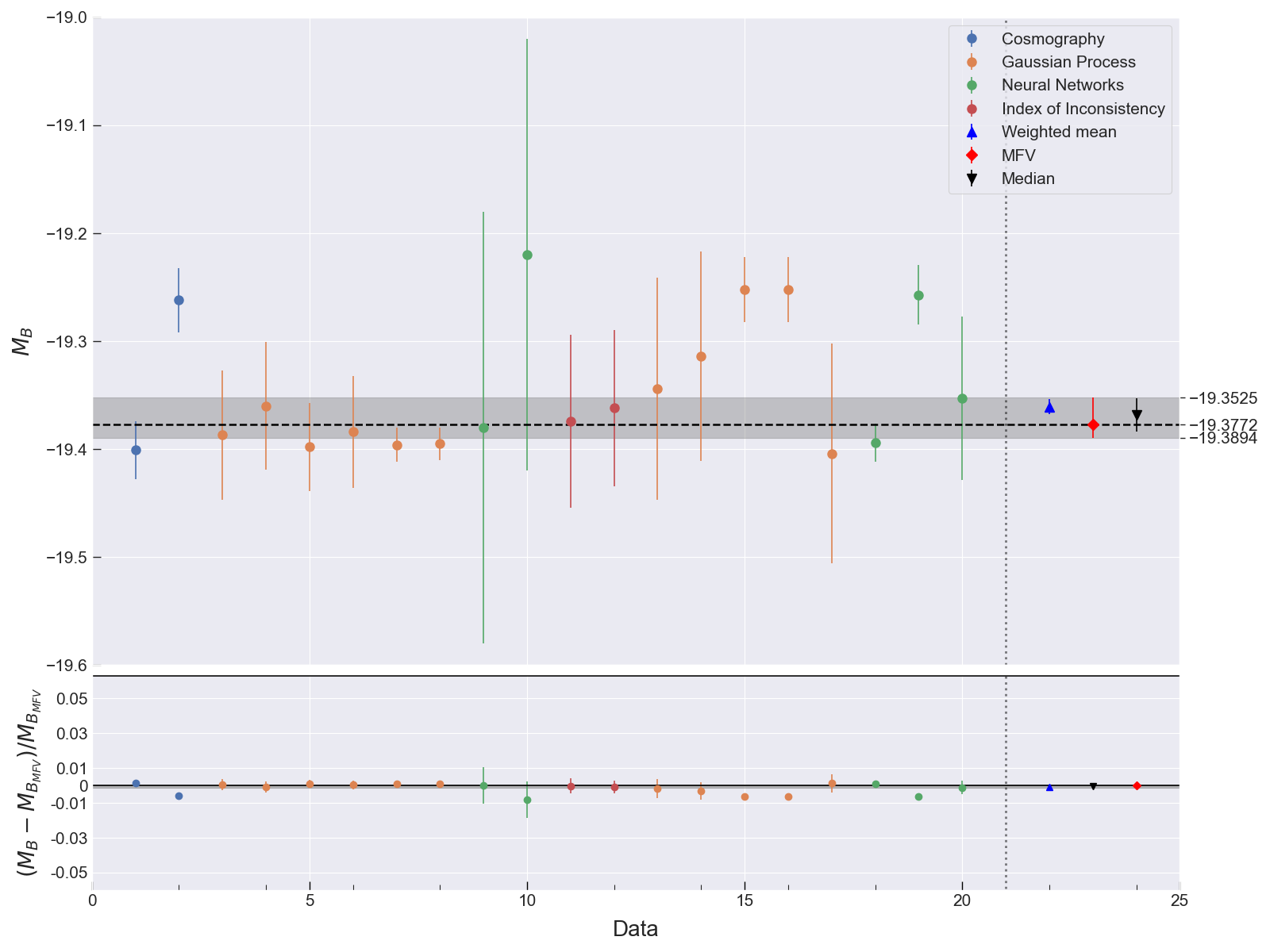}
	\caption{ (Top) $M_B$ as a function of the number of references. The points represent the $M_B$ data using different methods, whereas the triangle, diamond, and inverted triangle indicate the weighted mean, MFV, and median, respectively. Dark grey shadings display that the MFV uncertainties are at a confidence level of 1$\sigma$. (Bottom) The residuals of the fit or data are depicted at the bottom, showing their relationship with the MFV across different data points. [A colour version of this figure is available in the online version.]}
	\label{MB}
\end{figure}

Various efforts have been made to estimate $M_B$ from diverse combinations of cosmological observations \citep{Shah2024ApJS,Mukherjee2021MN,Mukherjee2021EPJC,Mukherjee2024JCAP,Dinda2023,Favale2023MN,Banerjee2023JCAP,Moresco2023arXiv,Gomez-Valent2022PRD}. 
Using these previous results and the bootstrap method, the holistic estimate of $M_B$ for all data is -19.3772 and the dihesion $\varepsilon=0.0291$.
Considering statistical bootstrap errors, the calculated 68.27 per cent confidence interval (CI) for all data is [-19.389, -19.352], while the 95.45 per cent confidence interval for all data is [-19.396, -19.319]. 
In Figure \ref{MB}, the distributions of $M_B$ are plotted as a function of reference number, showing the calculated results using the weighted mean, median, and MFV represented by different symbols. The residuals derived from the MFV are presented in the bottom panel. Similar outcomes were observed in the literature cited,  as noted by Mukherjee et al. \citep{Mukherjee2024JCAP}. This implies that error distributions are not Gaussian, thereby bolstering the credibility and rationality of the MFV estimate.

\subsection{Bayes based on $\Lambda$CDM}
By leveraging the advantages of the attention mechanism in extracting global information and the adaptive function approximation capabilities of KAN, this paper introduces the KLT-Net approach to estimate the distribution of $P(m\mid z)$, $\forall z \in \Bbb R^+$  and enhance the accuracy of pattern recognition in the Pantheon+ dataset $D=\{(z_i,m_i,\triangle m_i)\mid\forall i\in \{1,\ldots N\}\},z_i,m_i\in \Bbb R$ adopted from \citep{Shah2024ApJS}, through the integration of KAN, LSTM, and Transformer mechanisms. Then, we apply the Markov chain Monte Carlo (MCMC) method to estimate the best-fitting values and confidence intervals of the parameters in the $\Lambda$CDM model.

We employ the Bayesian approach for parameter estimation. 
The essence of Bayesian methodology is to characterize the probability distribution of parameters using the posterior distribution. The posterior distribution merges the prior distribution $P(\theta)$(initial understanding of the parameters) and the likelihood function $P(D\mid\theta)$ (the constraints of the data on the parameters). The Bayesian theorem can be expressed as:
\begin{eqnarray}
	P(\theta \mid D)=\frac{P(D\mid\theta)P(\theta)}{P(D)}.
\end{eqnarray}
In this scenario, the parameter $\theta=(H_0,\Omega_m)$ and the dataset $D$ refer to the observational data of supernovae SNIa, including redshift $z$, apparent magnitude $m(z)$, and its error $\sigma$. Given that the observation error is Gaussian distributed, the likelihood function can be calculated as:
\begin{eqnarray}
	L\varpropto exp(-\frac{1}{2}\chi^2(H_0,\Omega_m)),
\end{eqnarray}
where the $\chi^2$ function is represented as \citep{Wang2020ApJS,Zammit2024}
\begin{eqnarray}
	\chi^2(H_0,\Omega_m)= \sum^n_i(\frac{\mu_{SN}(z_i)-\mu_{th}(z_i;H_0,\Omega_m)}{\sigma_i})^2.
\end{eqnarray}
The uncertainty of $\mu$ for the function $\mu=m(z)-M_B$ can be evaluated using the error propagation formula. Asymmetrical uncertainties are considered here, with a preference for the maximum value for a conservative estimate \citep{Zhang2022,Barlow2003,Audi2017}.
Suppose that the prior distributions for $H_0$ and $\Omega_m$ follow uniform distributions, and the posterior distribution is obtained by multiplying the likelihood function and the prior distribution:
\begin{eqnarray}
	P(\theta \mid D)\varpropto{P(D\mid\theta)P(\theta)}.
\end{eqnarray}
After taking the logarithm, we obtain the posterior log distribution as:
\begin{eqnarray}
	{\rm log} P(\theta \mid D)={\rm log}{P(D\mid\theta)+{\rm log}P(\theta)}.
\end{eqnarray}
By substituting the likelihood function and prior distribution, we can obtain:
\begin{eqnarray}
	{\rm log} P(\theta \mid D)=-\frac{1}{2}\chi^2(H_0,\Omega_m)+{\rm log}P(\theta).
\end{eqnarray}
Due to the high dimensionality and complexity of the posterior distribution, we employ the MCMC method for sampling. MCMC generates a series of parameter samples through random walks, approximating the posterior distribution. Furthermore, MCMC realizes the estimation of the best fitting point and confidence interval by simulating the Markov chain in the parameter space. When the chain simulated by this method is long enough (reaching the preset upper limit of the number of samples) or meets the convergence requirements, it enters a stable state, then the chain terminates, and the sampling ends. This method ensures that the sampling points are concentrated near the best fitting point and the concerned area, thus avoiding the waste of time and resources.

\begin{figure}
	\centering
	\includegraphics[width=\textwidth, angle=0]{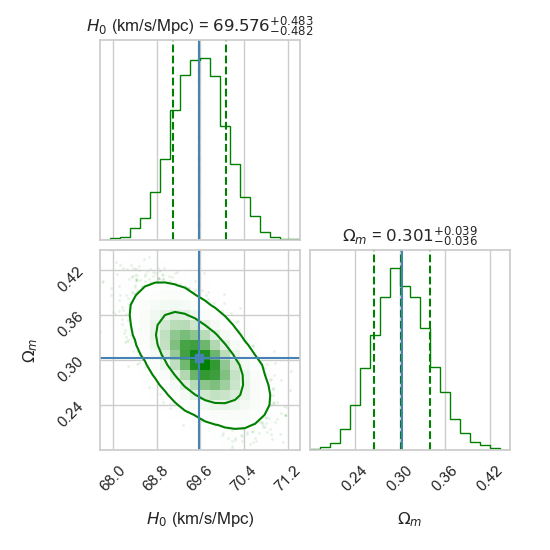}
	\caption{Comparison of cosmological parameter constraints derived from the KLT-Net predictions. In the contour diagram, the solid lines from inner to outer are the 1$\sigma$, 2$\sigma$ (CI), and represent the results after marginalization processing.}
	\label{corner}
\end{figure}

In Figure \ref{corner}, we present a corner plot of the two-dimensional joint confidence intervals for the cosmological parameters $H_0$ and $\Omega_{m0}$ in the $\Lambda$CDM model using the emcee \citep{emcee}.
By using the method described above, our program simulates a Markov chain in the free parameter space.
Then, we analyze the chain, remove the sampling points before burn-in, project the Markov chain onto the parameter space, and estimate the best fit parameters and confidence intervals by counting the number of times the Markov chain passes through different parameter values around a reference value.
The greater the frequency of occurrence, the higher the probability of being the optimal fit point. The most frequent point is the best fit point and the corresponding confidence interval is derived from this process. 

\begin{figure}
	\centering
	\includegraphics[width=\textwidth, angle=0]{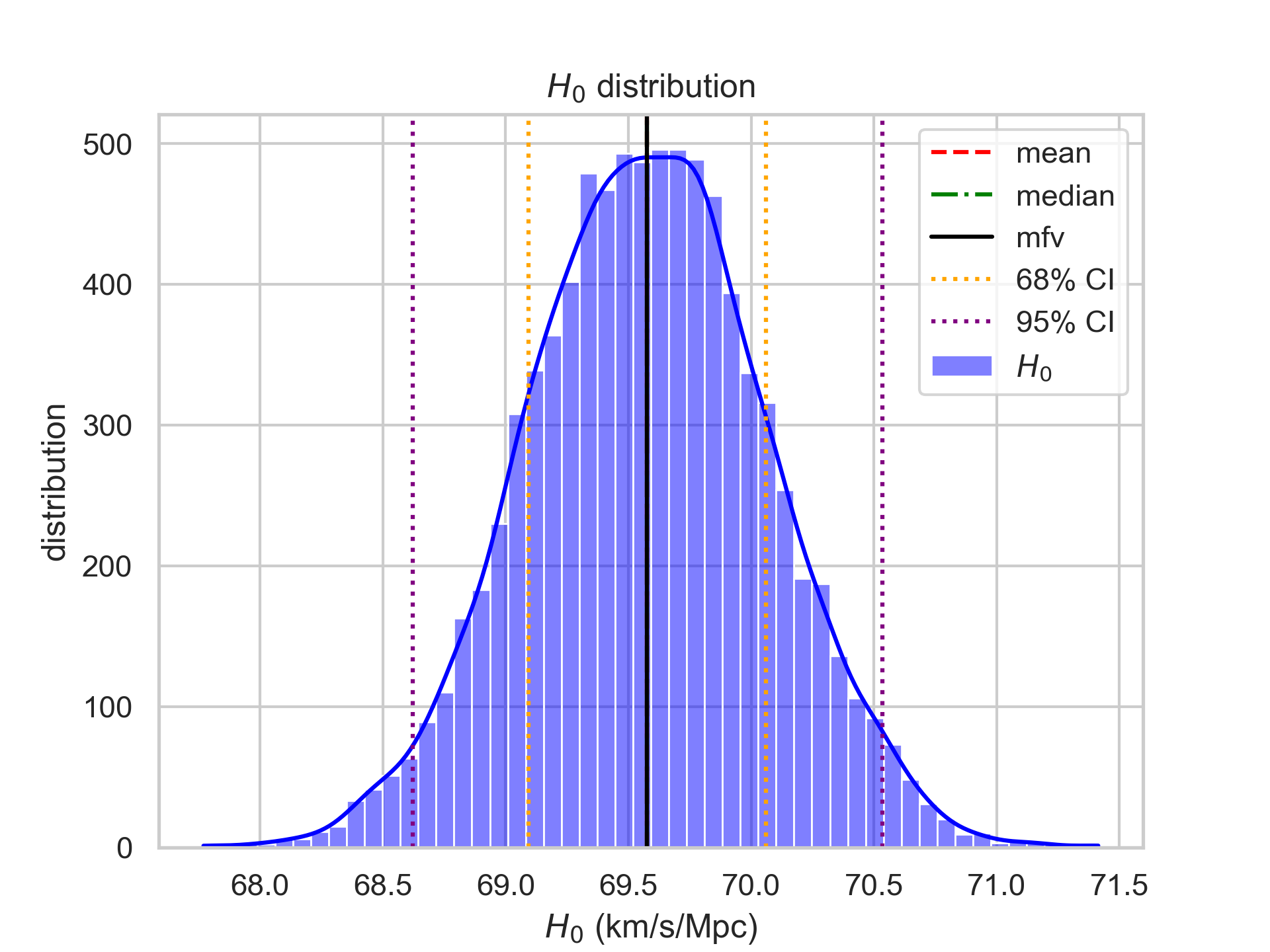}
	\caption{Distribution of $H_0$ obtained using MCMC vs. KLT-Net predictions.}
	\label{H0}
\end{figure}

\begin{figure}
	\centering
	\includegraphics[width=\textwidth, angle=0]{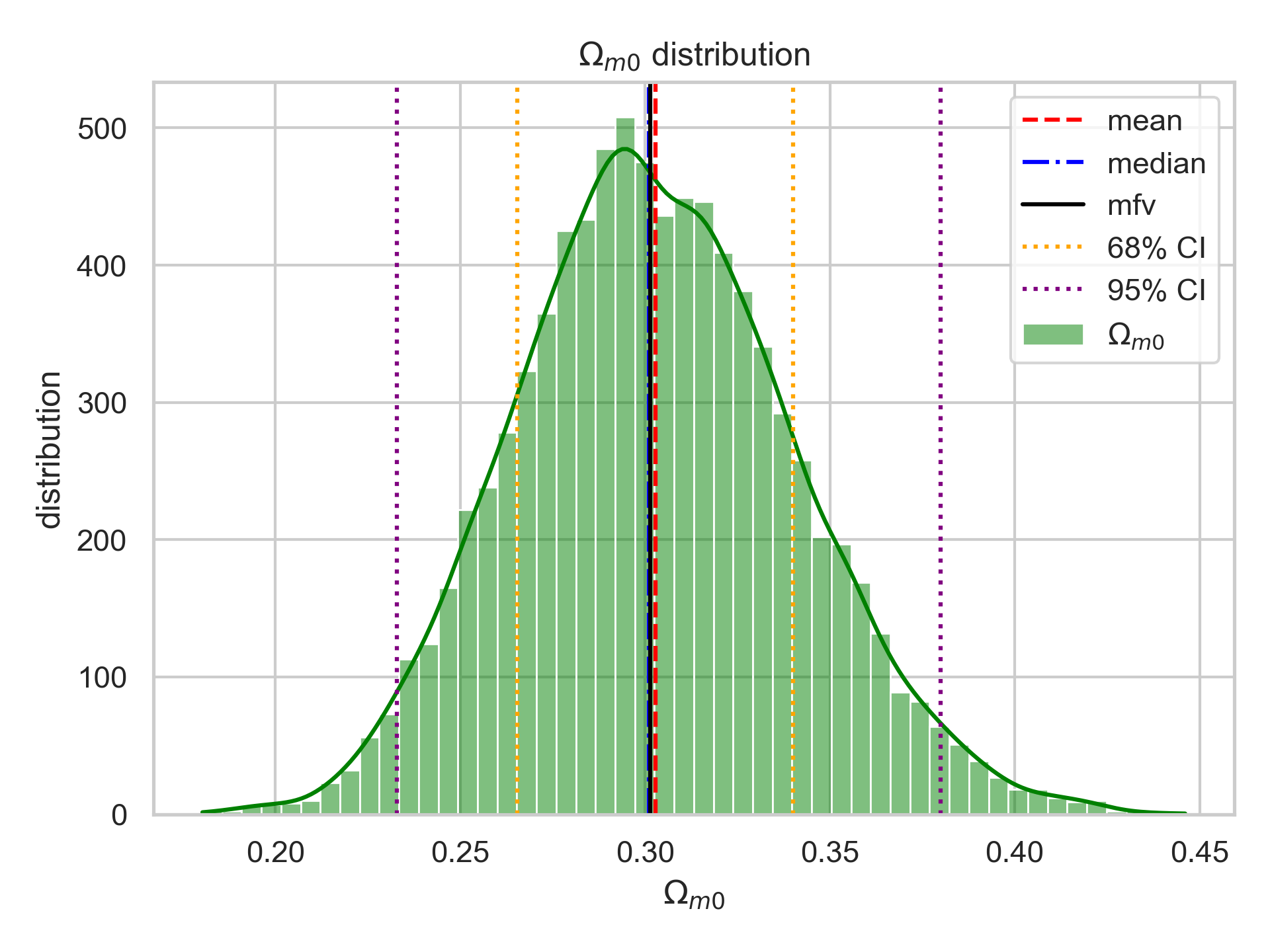}
	\caption{Distribution of $\Omega_{m0}$ obtained using MCMC vs. KLT-Net predictions.}
	\label{Omega}
\end{figure}

Based on the above methods, the minimum $\chi^2$ method yields $H_0=68.718\pm 0.350$ km/s/Mpc and $\Omega_{m0}=0.349\pm 0.025$ using the Pantheon data, while the holistic estimate for the KLT-Net prediction values is  $H_0=69.581\pm0.481$ km/s/Mpc and $\Omega_{m0}=0.300\pm 0.037$ along with the corresponding 1$\sigma$ uncertainty.  
Applying emcee, the MCMC method yields $H_0=68.716^{+0.338}_{-0.367}$ km/s/Mpc and $\Omega_{m0}=0.350^{+0.025}_{-0.025}$ using the Pantheon data, while the estimate for KLT-Net prediction values are $H_0=69.576^{+0.483}_{-0.482}$ km/s/Mpc and $\Omega_{m0}=0.301^{+0.039}_{-0.036}$ along with the corresponding 1$\sigma$ uncertainty.
This is consistent with the results that have been recently made public.
In addition to MFV, median, and mean, Figure \ref{H0} and \ref{Omega} also show the 68\% and 95\% CIs of $H_0$ and $\Omega_{m0}$. 

From these results, it can be seen that the results of $H_0$ are highly consistent and $\Omega_{m0}$ is different. This disparity is not only anticipated, but also reflects the essence of these methods and the characteristics of the different dataset. 
The core reasons are as follows: First, it is due to the sensitivity difference between the parameters and the dataset. 
Consequently, $\Omega_{m0}$ is primarily constrained by mid-to-high redshift data, where the dynamics of the universe are more sensitive to the matter density. At low redshift, the distance modulus is insensitive to $\Omega_{m0}$ (because the cosmological constant term is not dominant). Therefore, whether Pantheon or KLT-Net data are used, the estimation of  $H_0$ will be stable as long as the low red shift is consistent.
And $\Omega_{m0}$ is mainly determined by the data of middle and high redshift. Here, the geometry of the universe is determined by the competition between matter and dark energy. If the sample set contains more supernovae with high redshift and high accuracy, the constraint on $\Omega_{m0}$ is stronger.

Secondly, there are essential differences between the two datasets. The Pantheon sample comprises direct astronomical observations, whereas the KLT-Net outputs are model predictions derived from a trained network. The network's predictions might smooth over high-redshift features or inherit biases present in its training data. The large error and low central value of $\Omega_{m0}$ indicate that the prediction model may not fully capture the cosmological signal with high red shift in measrement data. This reflects the need for more high redshift observation data to optimize network prediction.

Finally, it may be the difference of statistical methods. The minimum $\chi^2$ method is sensitive to the peak value of likelihood surface to find a single best fitting point. In contrast, Bayesian statistics samples the entire posterior probability distribution, naturally accounting for parameter correlations and providing more robust uncertainty estimates. If there is correlation between $H_0$ and $\Omega_{m0}$ in posterior distribution, Bayesian method will give more robust uncertainty estimation. This motivates our use of the third method for verification.

On the other hand, we can also use the method of the Hessian ($\nabla^2f$) matrix to study the above problems\citep{Bishop2006,Bishop2023,Press2007,Hastie2009}. From the perspective of statistical learning and optimization theory, under the assumption of Gaussian error, minimizing the chi-square function $\chi^2$ is equivalent to maximizing the log-likelihood function ln$L$. Therefore, the minimum point of $\chi^2$ corresponds to the maximum point of ln$L$.  In order to address this problem, the finite difference algorithm is utilized, which demands the second-derivative or Hessian matrix, 
\newcommand{\hessian}[1]{\mathcal{H}\left(#1\right)}
\newcommand{\secondderiv}[3]{\frac{\partial^2 #1}{\partial #2 \partial #3}}
\begin{eqnarray}	
	\hessian{\chi^2} = 
	\begin{bmatrix}
		\secondderiv{\chi^2}{H_0}{H_0} & \secondderiv{\chi^2}{H_0}{\Omega_m} \\
		\secondderiv{\chi^2}{\Omega_m}{H_0} & \secondderiv{\chi^2}{\Omega_m}{\Omega_m} 
	\end{bmatrix}.
\end{eqnarray}

Using the finite difference method, we can calculate the elements of the Hessian matrix, such as
\begin{eqnarray}
	\frac{\partial^2 \chi^2}{\partial H^2_0} \approx \frac{\chi^2(H_0+\varDelta H_0,\Omega_m)-2\chi^2(H_0,\Omega_m)+\chi^2(H_0-\varDelta H_0,\Omega_m)}{(\varDelta H_0)^2}.
\end{eqnarray}
Similarly, other second derivatives and mixed derivatives can be calculated. Furthermore, the covariance matrix C is the inverse of the Hessian matrix $\frac{1}{2}\mathcal{H}$, and the error of parameters is the square root of the diagonal elements of the covariance matrix. 
The calculated $H_0$ for Pantheon data is $68.718\pm 0.352$ km/s/Mpc, while the calculated $H_0$ for KLT-Net predictions is $69.581\pm$ 0.481 km/s/Mpc  along with corresponding 1$\sigma$ uncertainty. The calculated $\Omega_{m0}$ for  Pantheon data is $0.349\pm0.025$, while the calculated $\Omega_{m0}$ for KLT-Net predictions is $0.300\pm$0.037, along with corresponding 1$\sigma$ uncertainty. 
These results are consistent with the outcomes from the above analysis.
It can be seen that the results of Hessian and $\chi^2$ methods are high consistent, which verifies the self-consistency of  statistical analysis. Bayesian method can capture the ``tailing" effect, which lead to the posterior mean deviating from the boundary, thus creating a slight gap with the classical method.

\section{Discussion}
Our findings underscore an essential differentiation between point-wise accuracy and training stability. 
Systematic deviations resulting from random weight initialization in cosmological studies have the potential to amplify, leading to notable biases in key parameters such as $H_0$ and $\Omega_{\rm m}$.
Consequently, although more structurally complex, the complete KLT-Net is better suited for Bayesian cosmological inference pipelines due to its efficient suppression of spurious fluctuations in reconstructed distance moduli $\mu(z)$.

Within this workflow, it is critical to ensure robustness against weight initialization. 
Excessive sensitivity to random seeds can result in undesirable scatter in parameter posterior distributions due to the direct impact of the reconstructed $\mu(z)$ on Bayesian inference of $H_0$ and $\Omega_{\rm m}$. 
No parametric cosmological model is utilized during network training by KLT-Net, emphasizing its model-independent reconstruction approach. 
However, the determination of $H_0$ and $\Omega_{\rm m}$ that follows is still dependent on the flat $\Lambda$CDM model. 
In the upcoming discussion, our goal is to lessen this dependency by integrating physical restrictions directly into the network framework.

There is room for enhancement in the current two-stage workflow to enhance physical consistency.
Recent advancements in physics-informed machine learning offer a promising approach to diminish the dependence on pre-established cosmological frameworks. 
The general framework for mechanism-guided deep learning was established by \cite{Zhang2023towards}, demonstrating the benefits of incorporating physical rules into neural networks. 
In accordance with \cite{Zeraatkar2026arXiv}, the Physics-Guided Transformer (PGT) was introduced. It incorporates heat-kernel bias into self-attention modules to enforce physical inductive biases. 
Outperforming conventional Physics-Informed Neural Networks (PINNs) by a considerable margin in the reconstruction of partial differential equations with limited data, this method's design philosophy can be readily adapted to cosmological distance reconstruction tasks.

The incorporation of physical priors in advanced scientific machine learning techniques has proven to significantly alleviate systematic biases in data-driven reconstructions \citep{Du2026JASTP,Li2026JGRA,Liu2026ESS}. 
Inspired by these successful paradigms in neighboring fields such as geophysics and atmospheric sciences, it is imperative to employ analogous physics-informed approaches (e.g., upholding the cosmological principle and constraints on luminosity evolution) for the reconstruction of the distance modulus in Type Ia supernovae.

The fusion of deep learning and basic physics is progressing from data reconstruction to the autonomous discovery of physical principles.
\cite{Tenachi2023ApJ} introduced the dimensionality-constrained Physics-Guided Symbolic Regression ($\Phi$-SO). This method extracts analytical physical equations from noisy astronomical data, ensuring adherence to unit consistency principles.
Furthermore, a framework was introduced by Bom et al. \cite{Bom2026arXiv} that utilizes large language models (LLMs) to engage in iterative physical reasoning aimed at exploring and enhancing dark energy equations of state. 
Future research could integrate the dimensional consistency of $\Phi$-SO with the reasoning capacity of LLMs to achieve end-to-end inference of dark energy equations of state or modified gravity theories, minimizing prior model assumptions.

\color{black}
\section{Conclusion}

In this study, the KLT-Net model is introduced to the field of astrophysical parameter estimation. Its key innovations include: (1) Theoretical inspiration: A differentiable network architecture is constructed using learnable spline basis functions, achieving parametric approximation of the multivariate-to-univariate function superposition described by the Kolmogorov-Arnold representation theorem. (2) Computational efficiency: A parameter sharing mechanism and gradient-guided adaptive mesh refinement are proposed, which solve the problem of high computational complexity of spline functions under high-dimensional input. (3) Physical interpretability: The path weights learned by the network directly map the contribution of input features (such as redshift and luminosity) to the output parameters, which are mutually verified with astrophysical prior knowledge.

Validation of the efficacy of these components is achieved through a systematic ablation study for the reconstruction of the cosmic distance ladder. 
The findings indicate that the KAN module markedly enhances distribution fitting capacity and improves training stability overall. 
Moreover, the LSTM module effectively captures extended sequential dependencies, thereby decreasing performance variability among various training iterations.
The Transformer module enhances model robustness by incorporating global context modeling, resulting in more cohesive reconstructions but with a minor decrease in point-wise prediction accuracy.
	
As shown in Table \ref{tab:model_results}, the KLT model achieves the lowest MSE (0.015727), demonstrating that the accuracy of cosmological parameter estimation critically depends on the reliability of the extended measurement dataset. On the other hand, using the MFV and bootstrap methods, we obtain the $M_B=-19.3772^{+0.0247}_{-0.0122}$ along with corresponding 1$\sigma$ uncertainty. Furthermore, based on the $\Lambda$CDM cosmological model, we use Bayesian methods to infer cosmological parameters by integrating SN data with theoretical models.
	
In summary, the ongoing challenge in astrophysics and cosmology research lies in the discrepancies of measurements for the Hubble constant $H_0$. In this article, we propose the KLT-Net method, which combines the advantages of KAN, LSTM, and Transformer mechanisms. 
By utilizing the KLT-Net procedure, the range of cosmological parameters can be estimated from the Pantheon dataset. 
Furthermore, we analyze the cosmological parameters of this model by restricting the best-fit values of the results via the MCMC method. The outcome of these calculations is $H_0=69.576^{+0.483}_{-0.482}$ km/s/Mpc and $\Omega_{m0}=0.301^{+0.039}_{-0.036}$ along with corresponding 1$\sigma$ uncertainty, which is validated by the Hessian matrix. In addition, we employ the MFV method to address the problem of data central estimate.

Overall, the complete KLT-Net demonstrates superior stability across various random initializations. For the purpose of ensuring consistent performance in repeated training sessions for cosmological parameter estimation, the suitable architecture is KLT-Net.
Based on this robustness, numerical simulations further show that the model enhances high-redshift generalization performance and parameter estimation efficiency.  By integrating the KLT-Net, MFV, and Bayesian statistics, we have established a robust framework for inferring cosmological parameters. This advancement offers a novel way to model extensive cosmological data in the future, laying the groundwork for astrophysical research in cosmology, especially in handling intricate datasets and high-dimensional parameter spaces.

\backmatter

\bmhead{Acknowledgments}

We thank Z.-W Han, B. Zhang and E. Feigelson for valuable discussions. This work was supported by the National Natural Science Foundation of China (Grant No. 11547041, 11403007, 11701135, 11673007), the Natural Science Foundation of Hebei Province (A2017403025, A2021403002).

\bibliography{mybib}


\end{document}